\documentstyle{amsppt}\TagsOnRight\nologo
\let\cal=\Cal
\def\Tr{\operatorname{Tr}}
\def\wt{\operatorname{wt}}
\newcount\refcount
\advance\refcount 1
\def\newref#1{\xdef#1{\the\refcount}\advance\refcount 1}
\newref\James
\newref\invariants
\newref\shadow
\newref\unitaryenum
\newref\shorlaflamme

\topmatter
\title Monotonicity of the quantum linear programming bound \endtitle
\author Eric M. Rains\endauthor
\affil AT\&T Research \endaffil
\address AT\&T Research, Room C290, 180 Park Ave.
         Florham Park, NJ 07932-0971, USA \endaddress
\email rains\@research.att.com \endemail
\date February 17, 1998 \enddate
\abstract
The most powerful technique known at present for bounding the size of
quantum codes of prescribed minimum distance is the quantum linear
programming bound.  Unlike the classical linear programming bound, it is
not immediately obvious that if the quantum linear programming constraints
are satisfiable for dimension $K$, that the constraints can be satisfied
for all lower dimensions.  We show that the quantum linear programming
bound {\it is} monotonic in this sense, and give an explicitly monotonic
reformulation.

\endabstract
\endtopmatter
\head Introduction\endhead
The most powerful technique known at present for bounding the size of
quantum codes of prescribed minimum distance is the quantum linear
programming bound:

\proclaim{Theorem (Quantum LP bound)}
If there exists a quantum code encoding $K$ states in $n$ qubits, with
minimum distance $d$, then there exist homogeneous polynomials $A(x,y)$,
$B(x,y)$, and $S(x,y)$ of degree $n$, satisfying the equations
$$
\align
B(x,y)&=A({x+3y\over 2},{x-y\over 2})\tag 1\\
S(x,y)&=A({x+3y\over 2},{y-x\over 2})\tag 2\\
A(1,0)&=K^2\tag 3\\
B(1,y)-{1\over K} A(1,y)&=O(y^d)\tag 4
\endalign
$$
and the inequalities
$$
\align
A(x,y)&\ge 0 \tag 5\\
B(x,y)-{1\over K}A(x,y)&\ge 0 \tag 6\\
S(x,y)&\ge 0, \tag 7
\endalign
$$
where $P(x,y)\ge 0$ means that the polynomial $P$ has nonnegative coefficients.
\endproclaim

\demo{Proof}
This is theorem 10 of \cite\shadow; see also \cite\shorlaflamme.  The
polynomials $A(x,y)$, $B(x,y)$, and $S(x,y)$ are the weight enumerator,
dual weight enumerator, and shadow enumerator, respectively, of the quantum
code.
\qed\enddemo

\demo{Remark}
In the sequel, we will use the standard notation $((n,K,d))$ to denote
a quantum code encoding $K$ states in $n$ qubits, with minimum distance
$d$.
\enddemo

It is clear that the existence of an $((n,K,d))$ code implies the existence
of an $((n,K',d))$ code for all $K'\le K$, which suggests that the same
should be true for the quantum LP bound, namely that if the quantum LP
constraints can be satisfied for $((n,K,d))$, then they can be satisfied
for $((n,K',d))$ for all $K'\le K$.  At first glance, this appears to be
false; after all, in the inequality \thetag{6}, decreasing $K$ actually
makes the inequality {\it harder} to satisfy.  This impression is
misleading, however; as we will see below, the quantum LP bound is indeed
monotonic in $K$.

\head 1. Random subcodes \endhead

The reason the quantum LP bound ``ought'' to be monotonic in $K$ is that if
${\cal Q}$ is an $((n,K,d))$ code, and $\hat{\cal Q}$ is a subcode of
${\cal Q}$ of dimension $K'$, then $\hat{\cal Q}$ is an $((n,K',d))$ code.
Of course, in general, it is impossible to deduce the weight enumerator of
$\hat{\cal Q}$ from the weight enumerator of ${\cal Q}$, so this is not
directly applicable to the LP bound.  However, if instead of picking a
specific subcode, we instead average over {\it all} subcodes of a given
dimension, the resulting average weight enumerator turns out to depend only
on the original weight enumerators.

Recall that if ${\cal Q}$ is an $((n,K,d))$ code, and $P_{\cal Q}$
is the orthogonal projection onto ${\cal Q}$, then the weight
enumerators $A_{\cal Q}(x,y)$ and $B_{\cal Q}(x,y)$ are
defined by
$$
\align
A_{\cal Q}(x,y)&=\sum_{e\in {\cal E}} \Tr(P_{\cal Q} e)^2 x^{n-\wt(e)}
y^{\wt(e)},\\
B_{\cal Q}(x,y)&=\sum_{e\in {\cal E}} \Tr(P_{\cal Q} e P_{\cal Q}
e)
x^{n-\wt(e)} y^{\wt(e)},
\endalign
$$
where ${\cal E}$ is the set of all tensor products of matrices from
the set $\{I,\sigma_x,\sigma_y,\sigma_x\}$, and $\wt(E)$ is the
number of nonidentity tensor factors in $E$.

Define
$$
\hat{A}_{\cal Q}(x,y)=E_{\hat{\cal Q}\subset {\cal Q}}
A_{\hat{\cal Q}}(x,y),
$$
and similarly for $\hat{B}_{\cal Q}(x,y)$, where the expectation is over
subcodes of dimension $K'$.  If we write $P_{\cal Q}=\Pi \Pi^\dagger$ for
some $2^n \times K$ matrix $\Pi$, then
$$
P_{\hat{\cal Q}} = \Pi P' \Pi^\dagger,
$$
for some $K\times K$ projection operator $P'$ with $\Tr(P')=K'$.  So
$$
\hat{A}_{\cal Q}(x,y)=E_{P'} \sum_{e\in {\cal E}} 
|\Tr(\Pi P' \Pi^\dagger e)|^2 x^{n-\wt(e)} y^{\wt(e)},
$$
and similarly for $\hat{B}_{\cal Q}$.  But
$$
\align
E_{P'} \Tr(\Pi P' \Pi^\dagger e)^2
&=
E_{U\in U(K)} \Tr(\Pi U P' U^\dagger \Pi^\dagger e)^2\\
&=
E_{U\in U(K)} \Tr(\Pi^\dagger e \Pi U P' U^\dagger)^2.
\endalign
$$
At this point, we can apply the following lemma:

\proclaim{Lemma 1}
Define functions
$$
\align
s_2(A)&={1\over 2}(\Tr(A)^2+\Tr(A^2))\\
s_{1^2}(A)&={1\over 2}(\Tr(A)^2-\Tr(A^2)).
\endalign
$$
For any $K\times K$ matrices $A$ and $B$,
$$
E_{U\in U(K)} s(A U B U^\dagger)
=
{s(A) s(B)\over s(I_K)},
$$
where $s$ is either $s_2$ or $s_{1^2}$.
\endproclaim

\demo{Proof}
This follows from the theory of zonal polynomials \cite\James.  For $A$ and
$B$ unitary, the relations follow from the fact that $s_2$ and $s_{1^2}$
are irreducible characters of the unitary group.  Since they are also
polynomial functions of $A$ and $B$, the relations must hold for arbitrary
matrices.
\qed\enddemo

In particular,
$$
\align
E_{U\in U(K)} \Tr(\Pi^\dagger e \Pi U P' U^\dagger)^2
&=
E_{U\in U(K)} s_2(\Pi^\dagger e \Pi U P' U^\dagger)
+s_{1^2}(\Pi^\dagger e \Pi U P' U^\dagger)\\
&=
 {{K'}^2+K'\over K^2+K} s_{2}(\Pi^\dagger e \Pi)+
 {{K'}^2-K'\over K^2-K} s_{1^2}(\Pi^\dagger e \Pi).
\endalign
$$
It follows that
$$
\hat{A}_{\cal Q}(x,y)
=
{K'(K' K-1)\over K^3-K} A_{\cal Q}(x,y)+
{K'(K-K')\over K^3-K} B_{\cal Q}(x,y).
$$
Similarly,
$$
\hat{B}_{\cal Q}(x,y)
=
{K'(K-K')\over K^3-K} A_{\cal Q}(x,y)+
{K'(K' K-1)\over K^3-K} B_{\cal Q}(x,y).
$$

In general, if $A(x,y)$ is a polynomial satisfying the quantum LP
constraints for $((n,K,d))$, then for any $K'\le K$, we can define
$$
\hat{A}(x,y) = 
{K'(K' K-1)\over K^3-K} A(x,y)+
{K'(K-K')\over K^3-K} B(x,y).
$$
The claim is that $\hat{A}$ satisfies the quantum LP constraints for
$K'$.  We have:
$$
\align
\hat{A} &= {{K'}^2\over K^2} A + {K' (K-K')\over K^3-K} (B-{1\over K} A)\\
\hat{B}-{1\over K'} \hat{A} &= {{K'}^2-1\over K^2-1} (B-{1\over K} A)\\
\hat{S} &=
{{K'}^2+K'\over K^2+K} \left({S(x,y)+S(-x,y)\over 2}\right)+
{{K'}^2-K'\over K^2-K} \left({S(x,y)-S(-x,y)\over 2}\right)
\endalign
$$
Since all of the constants appearing above are positive for $K'\le K$,
and $\hat{A}(1,0)={K'}^2$, the claim follows.  So we have proved:

\proclaim{Theorem 1}
The quantum linear programming bound is monotonic in $K$ for fixed $n$ and
$d$.
\endproclaim

\demo{Remark}
Similarly, the quantum LP bound for {\it pure} codes ($A(1,y)=1+O(y^d)$)
is monotonic in $K$, since the random subcode operator preserves 
purity.
\enddemo

We also obtain the following result of independent interest:

\proclaim{Theorem 2}
The average weight enumerator of a random $((n,K))$ quantum code is
$$
A(x,y) = {K (4^n K-2^n)\over 4^n-1} x^n+
{K (K-2^n)\over 4^n-1} (x+3y)^n.
$$
\endproclaim

\demo{Proof}
We have $A(x,y)=\hat{A}_{\cal H}$, where ${\cal H}$ is the trivial quantum
code consisting of the entire Hilbert space, with weight enumerator $4^n
x^n$.
\qed\enddemo

\head A reformulation \endhead

Lemma 1 suggests that we should be able to obtain a simpler formulation of
the quantum LP bound by considering the polynomials
$$
\align
C(x,y) &= {A(x,y)+B(x,y)\over K^2+K},\\
D(x,y) &= {A(x,y)-B(x,y)\over K^2-K}
\endalign
$$
(where $D(x,y)$ is only well-defined for $K>1$).  In particular, we
have the following result:

\proclaim{Lemma 2}
The polynomials $C$ and $D$ are preserved by the average subcode operator;
that is, $\hat{C}=C$ and $\hat{D}=D$.
\endproclaim

So, if we reformulate the quantum LP bound in terms of $C$ and $D$,
the result should be explicitly monotonic, in that a feasible solution for
$K$ will itself be a feasible solution for all smaller $K$.

\proclaim{Theorem 3}
If there exists an $((n,K,d))$ quantum code ($K>1$), then there exist
homogeneous polynomials $C(x,y)$ and $D(x,y)$, satisfying the equations
$$
\align
C(x,y)&=C({x+3y\over 2},{x-y\over 2})\tag 8\\
D(x,y)&=-D({x+3y\over 2},{x-y\over 2})\tag 9\\
C(1,0)&=1\tag 10\\
C(1,y)-D(1,y)&=O(y^d)\tag 11
\endalign
$$
and satisfying the inequalities
$$
\align
C(x,y)-{K-1\over 2K} (C(x,y)-D(x,y))&\ge 0\tag 12\\
C(x,y)-D(x,y)&\ge 0\tag 13\\
C({x+3y\over 2},{y-x\over 2})&\ge 0\tag 14\\
D({x+3y\over 2},{y-x\over 2})&\ge 0.\tag 15\\
\endalign
$$
\endproclaim

\demo{Proof}
We have
$$
\align
A(x,y) &= K^2 C(x,y) - {K^2-K\over 2} (C(x,y)-D(x,y))\\
B(x,y)-{1\over K}A(x,y) &= {K^2-1\over 2} (C(x,y)-D(x,y))\\
S(x,y) &= {K^2+K\over 2} C({x+3y\over 2},{y-x\over 2})+
          {K^2-K\over 2} D({x+3y\over 2},{y-x\over 2}).
\endalign
$$
Equations \thetag{8} and \thetag{9} are clearly equivalent to \thetag{1},
while \thetag{10} and \thetag{11} are together equivalent to \thetag{3} and
\thetag{4}.  Similarly, the inequalities \thetag{12} and \thetag{13} are
equivalent to \thetag{5} and \thetag{6} respectively.

For \thetag{14} and \thetag{15}, it suffices to note that \thetag{8} and
\thetag{9} imply
$$
\align
C({x+3y\over 2},{y-x\over 2})&=C({-x+3y\over 2},{y+x\over 2})\\
D({x+3y\over 2},{y-x\over 2})&=-D({-x+3y\over 2},{y+x\over 2})
\endalign
$$
It follows that the two terms in the expression for $S(x,y)$ have disjoint
support.  So \thetag{7} becomes \thetag{14} and \thetag{15}.
\qed\enddemo

Theorem 1 is an obvious corollary; $K$ appears only in \thetag{12},
and decreasing $K$ in that equation only makes the constraint easier
to satisfy.  For pure codes, the additional constraint $C(1,y)=1+O(y^d)$
holds, and again monotonicity is obvious.

It should also be noted that this theorem carries over readily to nonbinary
codes (from the inequalities in \cite\invariants and \cite\unitaryenum); in
particular, the quantum LP bound is monotonic for larger alphabet codes
as well.

\enddocument
\bye